\documentclass[sigconf]{acmart}

\usepackage{booktabs} 
\usepackage{multirow}
\usepackage{subfigure}
\usepackage{verbatim}
\usepackage{CJKutf8}
\usepackage{flushend}

\usepackage{makecell}
\usepackage[utf8]{inputenc}

\copyrightyear{2022}
\acmYear{2022}
\setcopyright{rightsretained}
\acmConference[SIGIR '22]{Proceedings of the 45th International ACM SIGIR Conference on Research and Development in Information Retrieval
}{July 11--15, 2022}{Madrid, Spain}
\acmBooktitle{Proceedings of the 45th International ACM SIGIR Conference on Research and Development in Information Retrieval
 (SIGIR '22), July 11--15, 2022, Madrid, Spain}
\acmPrice{15.00}

\begin{document}

\title{MM-Rec: Multimodal News Recommendation}


\author{Chuhan Wu$^1$, Fangzhao Wu$^2$, Tao Qi$^1$, Yongfeng Huang$^1$}

\affiliation{%
  \institution{$^1$Department of Electronic Engineering, Tsinghua University, Beijing 100084 \\ $^2$Microsoft Research Asia, Beijing 100080, China}
} 
\email{{wuchuhan15,wufangzhao,taoqi.qt}@gmail.com,yfhuang@tsinghua.edu.cn}






\begin{abstract}

News representation is critical for news recommendation.
Most existing news representation methods learn news representations only from news texts while ignoring the visual information in news like images.
In fact, users may click news not only due to the interest in news titles but also the attraction of news images.
Thus, images are useful for representing news and predicting news clicks. 
Pretrained visiolinguistic models are powerful in multi-modal  understanding, which can empower news representation learning from both textual and visual contents.
In this paper, we propose a multimodal news recommendation method that can incorporate both textual and visual information of news to learn multimodal news representations.
We first extract region-of-interests (ROIs) from news images via object detection. 
We then use a pre-trained visiolinguistic model to encode both news texts and image ROIs and model their inherent relatedness using co-attentional Transformers.
In addition, we propose a crossmodal candidate-aware attention network to select relevant historical clicked news for the accurate modeling of user interest in candidate news.
Experiments validate that incorporating multimodal news information can effectively improve news recommendation.
\end{abstract}

%
%


\maketitle

\section{Introduction}

\begin{figure}[!t]
  \centering
    \includegraphics[width=0.9\linewidth]{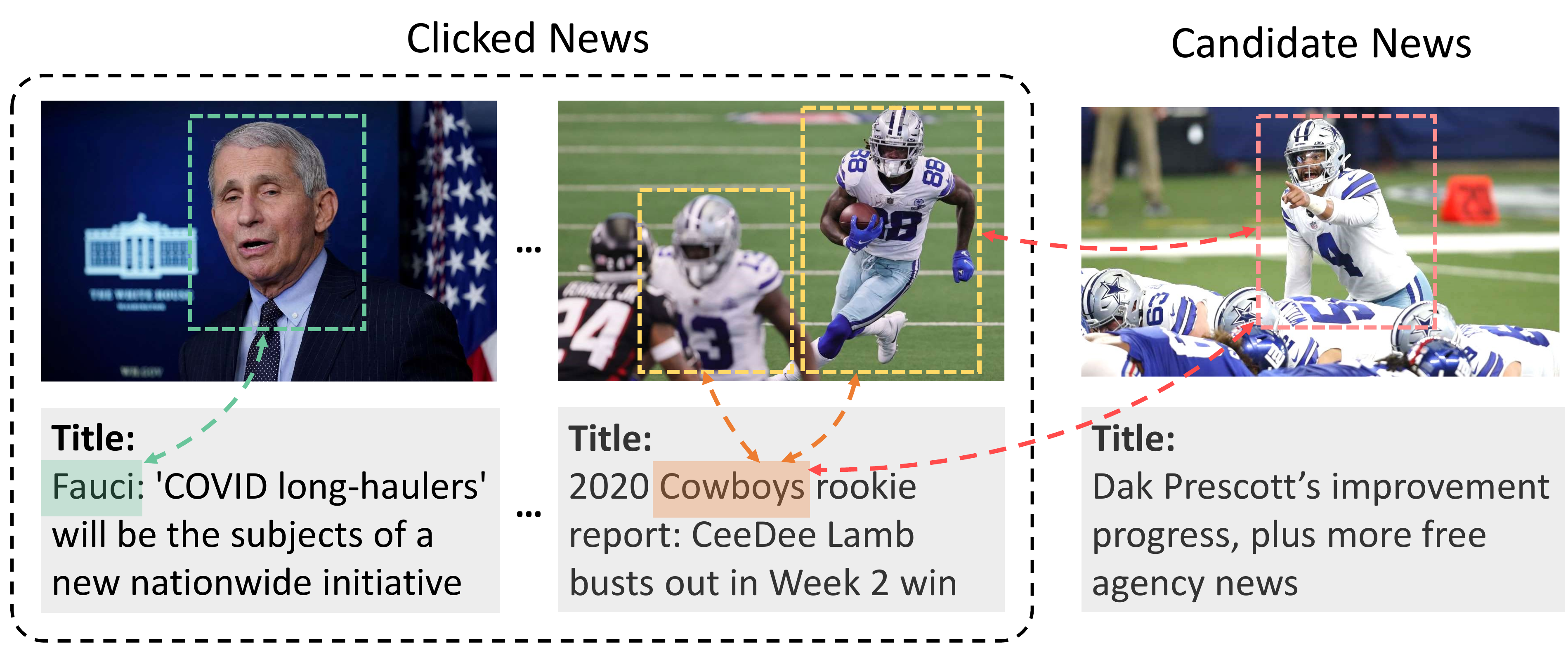}

  \caption{News with images for news recommendation.}
  \label{fig.example}\vspace{-0.1in}

\end{figure}

News representation is critical for news recommendation~\cite{wu2020mind}.
Most existing news representation methods learn news representations merely from news texts~\cite{okura2017embedding,wang2018dkn,wu2019,an2019neural,wu2020user,ge2020graph,hu2020graph,wang2020fine,tian2021joint,qi2021hierec,qi2021pp,zhang2021unbert,zhang2021amm,wu2021personalized}.
For example, \citet{okura2017embedding} used autoencoders to learn news representations from  news content.
\citet{wu2019npa} used CNN and personalized attention network  to learn news representations from news titles.
\cite{wu2019nrms} used multi-head self-attention networks to model news from titles.
In fact, besides the titles, many news websites also use images to better attract users' clicks~\cite{lommatzsch2018newsreel}, as shown in Fig.~\ref{fig.example}.
Users may click news to read not only because of the interest in the content of news title, but also due to the fascination of news images~\cite{bednarek2012value,jeong2020news,xun2021we}.
For example, in Fig.~\ref{fig.example} the image of the second clicked news shows a highlight moment in an NFL game, which may be attractive for users interested in football.
The visual information of news images can provide rich  information for news content understanding and future behavior prediction.

In this paper we study how to incorporate visual news information to enhance news recommendation.
Our work is motivated by the following observations.
First, news titles and images usually have some relatedness in describing news content and attracting clicks.
For example, in the second news of Fig.~\ref{fig.example}, the word ``Cowboys'' in news title is related to the players shown in the news image.
Modeling their relatedness can help better model news and infer user interest for news recommendation.
Second, a user may have multiple interests, and a candidate news may only be related to a specific interest encoded in part of clicked news. 
For example, in Fig.~\ref{fig.example} the candidate news is only related to the second clicked news.
Thus, modeling the relevance between clicked news and candidate news can help predict users' specific interest in a candidate news.
Moreover, candidate news may have crossmodal relatedness with clicked news.
In Fig.~\ref{fig.example} the image of candidate news is related to both the image and title of the second clicked news, because both images show the same football team and its name is mentioned by the title of the second clicked news.
Modeling the crossmodal relations between candidate news and clicked news can help measure their relevance accurately.

\begin{figure*}[!t]
  \centering
    \includegraphics[width=0.8\linewidth]{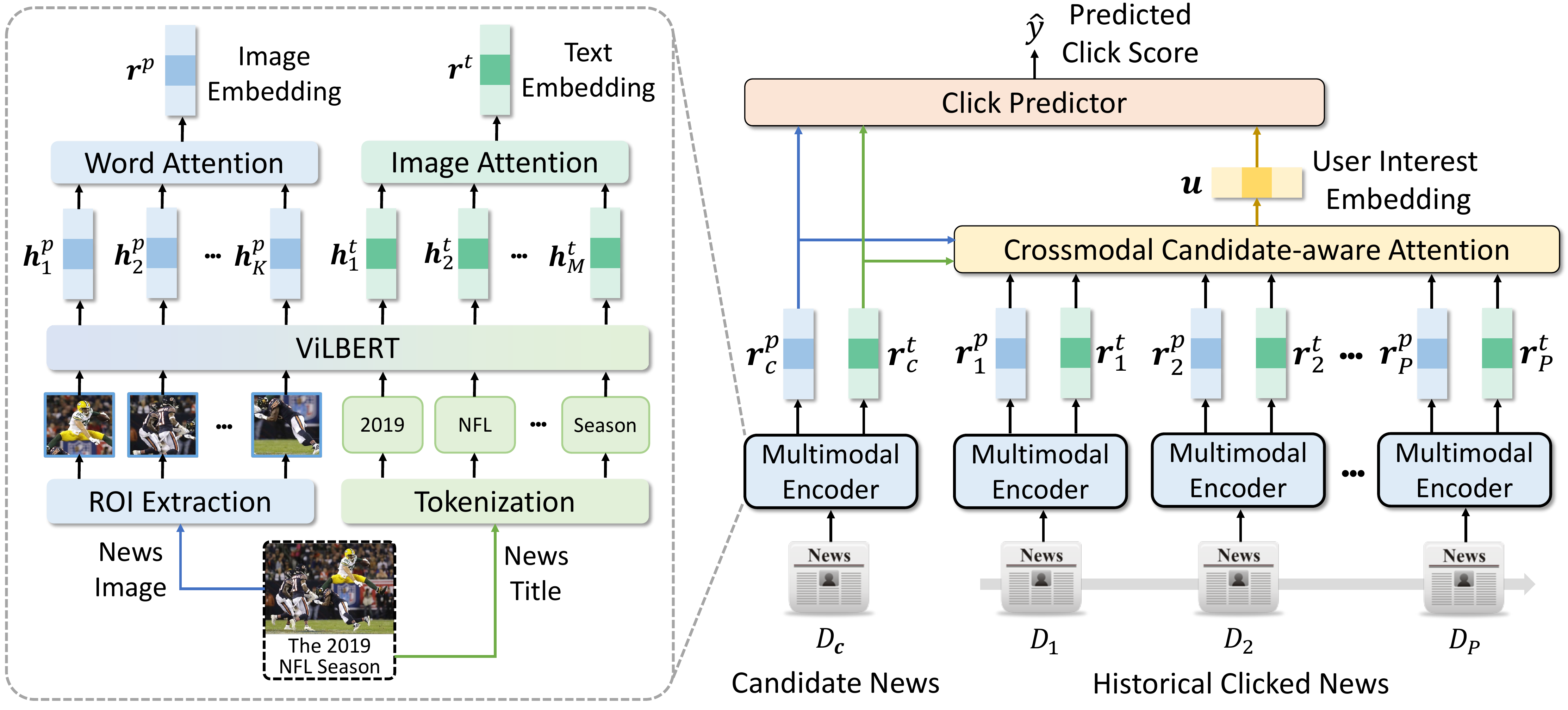}
\vspace{-0.06in}
  \caption{The model architecture of \textit{MM-Rec}.}
\vspace{-0.1in}
  \label{fig.model}
\end{figure*}

In this paper, we present a multimodal news recommendation method named MM-Rec, which  leverages both textual and visual news information for news recommendation.
In our approach, we first extract region-of-interests (ROIs) of news images via a pre-trained Mask R-CNN model~\cite{he2017mask} for object detection.
Then we use a pre-trained visiolinguistic model~\cite{lu2019vilbert} to encode both news texts and  news image ROIs and model their inherent crossmodal relatedness via co-attentional Transformers to learn accurate multimodal news representations.
In addition, we propose a crossmodal candidate-aware attention network to select relevant clicked news for user modeling by evaluating the crossmodal relevance between candidate news and clicked news, which can help  model users' specific interest in candidate news.
Experiments on real-world dataset show that incorporating multimodal news information can  effectively improve news recommendation.

\section{MM-Rec}
Next, we introduce our \textit{MM-Rec} method that uses both textual and visual news information for news recommendation, as shown in Fig.~\ref{fig.model}.
We first introduce its multimodal news encoder that learns multimodal news representations from texts and images, and then introduce how to make  news recommendations based on multimodal news representations.

\subsection{Multimodal News Encoder}\label{sec:Model}

On news websites, many news have images displayed together with titles, as shown in Fig.~\ref{fig.example}.
Users may click news not only due to their interests in news title, but also because of the attraction of news images~\cite{bednarek2012value}.
Thus, modeling  visual news content like images is important for  news representation.
Since different regions in news image may have different informativeness for news modeling, we use a Mask-RCNN~\cite{he2017mask} model pre-trained in an object detection task to extract ROIs of news images.
We further use a ResNet-50~\cite{he2016deep} model to extract features of ROIs, which forms a feature sequence $[\mathbf{e}^p_1,\mathbf{e}^p_2,...,\mathbf{e}^p_K]$, where $K$ is the number of ROIs. 
To model textual content, following previous works~\cite{wang2018dkn,wu2019nrms} we take news titles as model input.
We tokenize a news title into a word sequence $[w^t_1,w^t_2,...,w^t_M]$, where $M$ is the number of words.

An intuitive way is modeling news texts and images with separate models.
However, the title and image  of the same news usually have some relations.
For instance, in the first news of Fig.~\ref{fig.example}, the word ``Fauci'' in the news title is related to his photo.
Capturing the relatedness between news titles and images can help better understand their content and infer user interests.
Visiolinguistic models are effective in modeling the crossmodal relations between texts and images~\cite{lu2019vilbert,su2019vl,tan2019lxmert,li2020unicoder}.
Thus, we apply the pre-trained ViLBERT model~\cite{lu2019vilbert} to capture the inherent relatedness between news title and image when learning representations of them.
The inputs of ViLBERT are the ROI and word sequences.
It first models the contexts of words via several Transformers~\cite{vaswani2017attention}, and then use several co-attentional Transformers~\cite{lu2016hierarchical} to capture the crossmodal interactions between the image and title.
The output is a hidden ROI representation sequence $\mathbf{H}^p=[\mathbf{h}^p_1,\mathbf{h}^p_2,...,\mathbf{h}^p_K]$ and a hidden word representation sequence 
$\mathbf{H}^t=[\mathbf{h}^t_1,\mathbf{h}^t_2,...,\mathbf{h}^t_M]$.
Then we apply a word attention network to learn  title representations and an image attention network to learn image representations.
The attention weights of words in news title are computed as follows:
\begin{equation}
\mathbf{a}^t=\text{softmax}[(\mathbf{W}^t\mathbf{H}^t)^\top\mathbf{q}^t],
\end{equation}
where $\mathbf{q}^t$ is an attention query vector and $\mathbf{W}^t$ is a parameter matrix.
The final representation of news title is the summation of the hidden word representations weighted by their attention weights, i.e.,
$
\mathbf{r}^t=\mathbf{H}^t \times \mathbf{a}^t.
$
The attention weights of ROIs are computed in a similar way as follows:
\begin{equation}
\mathbf{a}^p=\text{softmax}[(\mathbf{W}^p\mathbf{H}^p)^\top\mathbf{q}^p],
\end{equation}
where $\mathbf{q}^p$  and $\mathbf{W}^p$ are parameters.
The final representation of news image is the summation of hidden ROI representations weighted by their attention weights, i.e.,
$
\mathbf{r}^p=\mathbf{H}^p \times \mathbf{a}^p.
$

\subsection{Multimodal News Recommendation}

Then we introduce how to make news recommendations using the multimodal news representations.
Since news recommendation usually relies on the relevance  between candidate news articles and users' personal interest to rank candidate news for a target user, we first introduce our user interest modeling method. 
Following many prior works~\cite{okura2017embedding,wu2019nrms}, we model users' interest in news from the representations of their previously clicked news.
We use the multimodal news encoder to learn the text and image representations of previously clicked news from their titles and images, which are denoted as $\mathbf{R}^t=[\mathbf{r}^t_1,\mathbf{r}^t_2,...,\mathbf{r}^t_P]$ and $\mathbf{R}^p=[\mathbf{r}^p_1,\mathbf{r}^p_2,...,\mathbf{r}^p_P]$, where $P$ is the number of clicked news.
However, not all clicked news are informative for inferring user interests on a candidate news, because it may be relevant to a few clicked news only.
For example, the candidate news in Fig.~\ref{fig.example} is only related to the second clicked news.
Thus, selecting clicked news according to their relevance to candidate news in user modeling may help accurately match candidate news with user interest.
In addition, candidate news may have some crossmodal relations with the images and titles of clicked news.
For example, in Fig.~\ref{fig.example} the players in the candidate news image are related to the players in the images of the second clicked news and the word ``Cowboys'' in its title.
Motivated by these observations, we propose a crossmodal candidate-aware attention network to measure the crossmodel relevance between clicked news and candidate news for  better modeling user interests in candidate news. 
We denote the image and text representations of a candidate news $D_c$ as $\mathbf{r}^t_c$ and $\mathbf{r}^p_c$, respectively.
We compute the text-text attention weights for clicked news that represent their text-text relevance to candidate news as follows:
\begin{equation}
\mathbf{a}^{t,t}=\text{softmax}(\mathbf{R}^t\times\mathbf{r}^t_c).
\end{equation}
In a similar way, we compute the text-image, image-text and image-image attention weights of clicked news as
$\mathbf{a}^{t,p}=\text{softmax}(\mathbf{R}^p\times\mathbf{r}^t_c),$ $\mathbf{a}^{p,t}=\text{softmax}(\mathbf{R}^t\times\mathbf{r}^p_c),$ and $\mathbf{a}^{p,p}=\text{softmax}(\mathbf{R}^p\times\mathbf{r}^p_c).$
The unified user embedding $\mathbf{u}$ is computed as:
\begin{equation}
\mathbf{u}=\mathbf{R}^p\times (\mathbf{a}^{t,p}+\mathbf{a}^{p,p})+\mathbf{R}^t\times (\mathbf{a}^{t,t}+\mathbf{a}^{p,t}).
\end{equation}

In our method, the news click score for ranking is derived from the multimodal representations $\mathbf{r}^t_c$ and $\mathbf{r}^p_c$ of candidate news and the user representation $\mathbf{u}$.
Motivated by~\cite{okura2017embedding}, the  click score $\hat{y}$ is predicted by $\hat{y} =  \mathbf{r}^t_c\times\mathbf{u}+\mathbf{r}^p_c\times\mathbf{u}$.
Following~\cite{wu2019nrms} we use negative sampling to build labeled samples from news click logs for model training and use cross-entropy as the loss function.

\section{Experiments}\label{sec:Experiments}

\subsection{Datasets and Experimental Settings}

Since there is no high-quality news recommendation dataset that contains multimodal information, we constructed one based on the logs collected from a commercial news website during three weeks (from Feb. 25, 2020 to Mar. 16, 2020).
Logs in the first week were used to construct user histories and the rest sessions were used to form click and non-click samples.
We sorted sessions by time and used the first 1M sessions for training, the next 100K sessions for validation and the rest for test.
 Table~\ref{dataset} shows the dataset statistics.

\begin{table}[h]

\centering
\caption{Statistics of our dataset.}
\label{dataset} 
\resizebox{0.48\textwidth}{!}{
\begin{tabular}{|l|r|l|r|}
\hline
\textbf{\# users}       & 536,289  & \textbf{\# clicked samples}     & 1,887,697     \\ 
\textbf{\# news}        & 152,723    & \textbf{\# non-clicked samples}     & 51,642,426       \\ 
\textbf{\# sessions} & 1,200,000 & \textbf{CTR of news w/ images}     & 0.0384     \\ 
\textbf{\# news w/ images} & 111,312 & \textbf{CTR of news w/o images}     & 0.0353 \\ \hline
\end{tabular}
}

\end{table}

In our experiments, we finetuned the last three layers of ViLBERT. 
We used Adam~\cite{kingma2014adam} as the optimizer (lr=1e-5).
The batch size was 32.
We tuned hyperparameters on the validation set.
We repeated each experiment 5 times and reported the average AUC, MRR, NDCG@5 and NDCG@10 scores.

\subsection{Performance Comparison}

We compare the proposed \textit{MM-Rec} method with many baseline methods, including:
(1)  \textit{EBNR}~\cite{okura2017embedding}, learning news embeddings via autoencoder and user embeddings with a GRU network;
(2) \textit{DKN}~\cite{wang2018dkn}, learning news representations via a knowledge-aware CNN model; 
(3) \textit{DAN}~\cite{zhu2019dan}, learning news representations with two parallel CNN networks from news title and entities;
(4) \textit{NAML}~\cite{wu2019}, using attentive multi-view learning to learn news representations;
(5) \textit{NRMS}~\cite{wu2019nrms}, using multi-head self-attention networks to learn news representations;
(6) \textit{GERL}~\cite{ge2020graph}, a graph enhanced news recommendation method;
(7) \textit{FIM}~\cite{wang2020fine}, a fine-grained matching method with 3-D CNN for news recommendation.
(8) \textit{PLM-NR}~\cite{wang2020fine}, a pre-trained language empowered approach for news recommendation.
We use BERT~\cite{wu2021empowering} as the news model.
In these methods, only news texts are considered.
The results are summarized in Table~\ref{table.result}.
It shows that our \textit{MM-Rec} approach that considers visual information of news outperforms other methods based on textual content only.
T-test results further validate the significance of improvement ($p<0.01$).
This is because users usually click news articles not only based on their interest in news texts, but also the attraction of news images.
Thus, the visual information of news images can enrich news representations for recommendation.
Our \textit{MM-Rec} method can incorporate both textual and visual news information into news representation learning and meanwhile model their inherent relatedness for better news content understanding, while in existing news recommendation methods the image-related information is ignored.
In addition, our approach can model the crossmodal relatedness between clicked news and candidate news for more accurate interest matching, which can yield better performance.

\begin{table}[!t]
	\centering
	\caption{Performance comparison of different methods.}\label{table.result} 
\resizebox{1.0\linewidth}{!}{
\begin{tabular}{ccccc}
\Xhline{1.5pt}
    \textbf{Methods}         & \textbf{AUC}             & \textbf{MRR}             & \textbf{NDCG@5}          & \textbf{NDCG@10}         \\ \hline
EBNR          & 60.34$\pm$0.29 & 20.79$\pm$0.25 & 22.43$\pm$0.26 & 30.76$\pm$0.23       \\ 
DKN          & 60.18$\pm$0.24 & 20.56$\pm$0.22 & 22.24$\pm$0.20 & 30.53$\pm$0.18        \\  
DAN          & 61.03$\pm$0.22 & 21.69$\pm$0.19 & 23.12$\pm$0.23 & 31.48$\pm$0.20 \\
NAML         & 61.55$\pm$0.18 & 22.13$\pm$0.16 & 23.57$\pm$0.17 & 31.92$\pm$0.17 \\ 
NRMS         & 62.01$\pm$0.13 & 22.68$\pm$0.15 & 24.08$\pm$0.15 & 32.38$\pm$0.15 \\  
GERL         & 62.21$\pm$0.17 & 22.82$\pm$0.16 & 24.36$\pm$0.18 & 32.55$\pm$0.19 \\ 
FIM         & 62.18$\pm$0.15 & 22.79$\pm$0.14 & 24.35$\pm$0.13 & 32.52$\pm$0.16 \\ 
PLM-NR         & 63.67$\pm$0.10 & 24.17$\pm$0.09 & 25.42$\pm$0.11 & 33.31$\pm$0.12 \\ \hline
MM-Rec          & \textbf{64.96}$\pm$0.12 & \textbf{25.22}$\pm$0.11 & \textbf{26.67}$\pm$0.12 & \textbf{34.23}$\pm$0.10 \\ \Xhline{1.5pt}
\end{tabular}
} 

\end{table}

\begin{figure}[!t]
	\centering
	\includegraphics[width=0.4\textwidth]{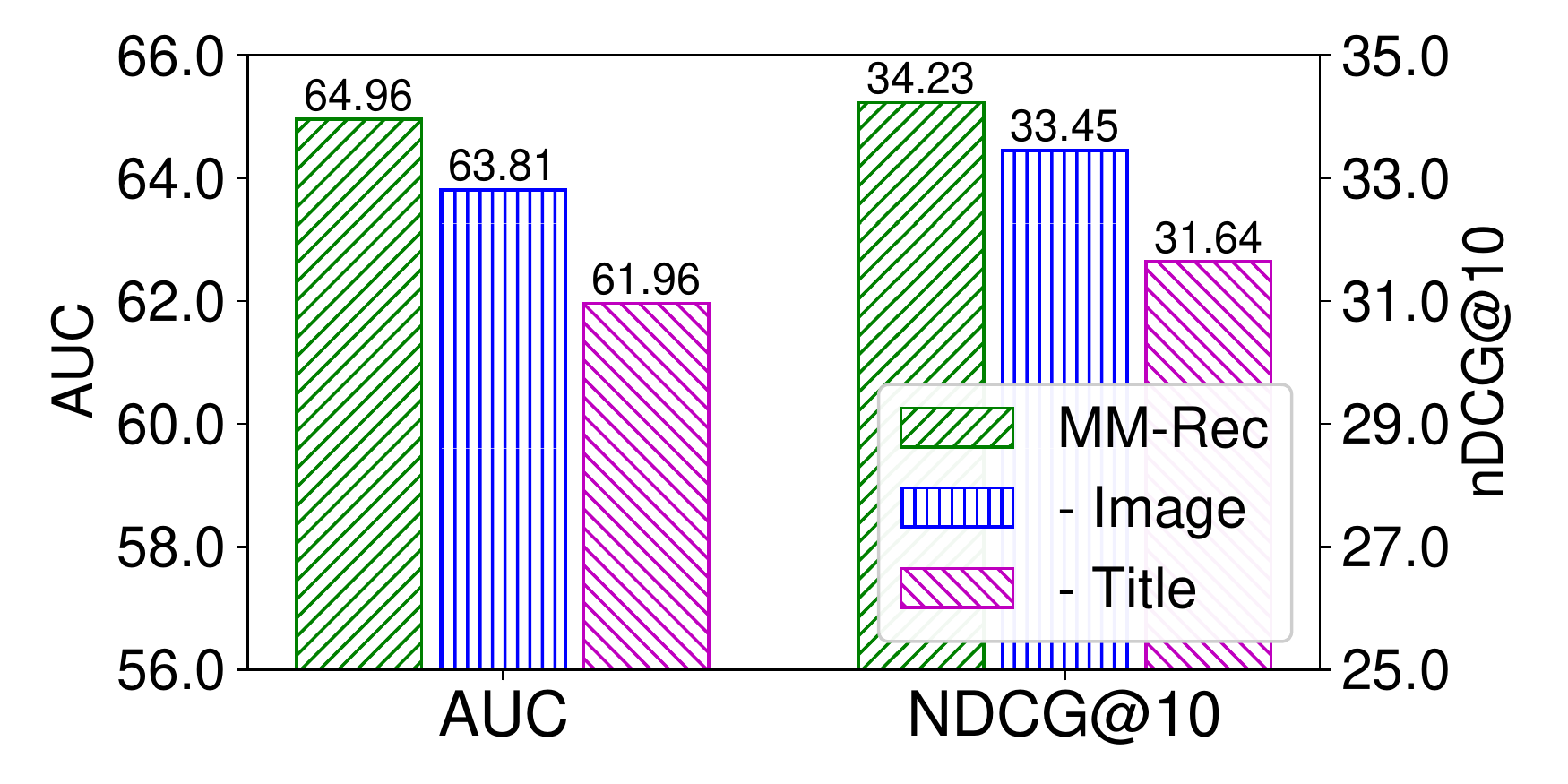}
\vspace{-0.05in}
\caption{Effectiveness of multimodal information.}\label{fig.view} 
\vspace{-0.15in}
\end{figure}

\subsection{Ablation Study}

Next, we study the effectiveness of multimodal information for news representation.
We compare \textit{MM-Rec} with its two variants with images or titles only.
The results are shown in Fig.~\ref{fig.view}. 
We find that both news title and image are useful for learning news representations for recommendation.
It shows that both textual and visual information of news are highly useful for understanding news content and inferring user interest.
We also have an interesting finding that the performance of \textit{MM-Rec} without image information is also slightly better than the results of PLM-NR baseline in Table~\ref{table.result}.
This is because the ViLBERT model is pretrained on multi-modal data, which can leverage visual signals to enhance text understanding.
In addition, incorporating multimodal news information can further improve the recommendation performance, which shows that incorporating multimodal news information can help learn accurate news representations.

\begin{figure}[!t]
	\centering
	\includegraphics[width=0.4\textwidth]{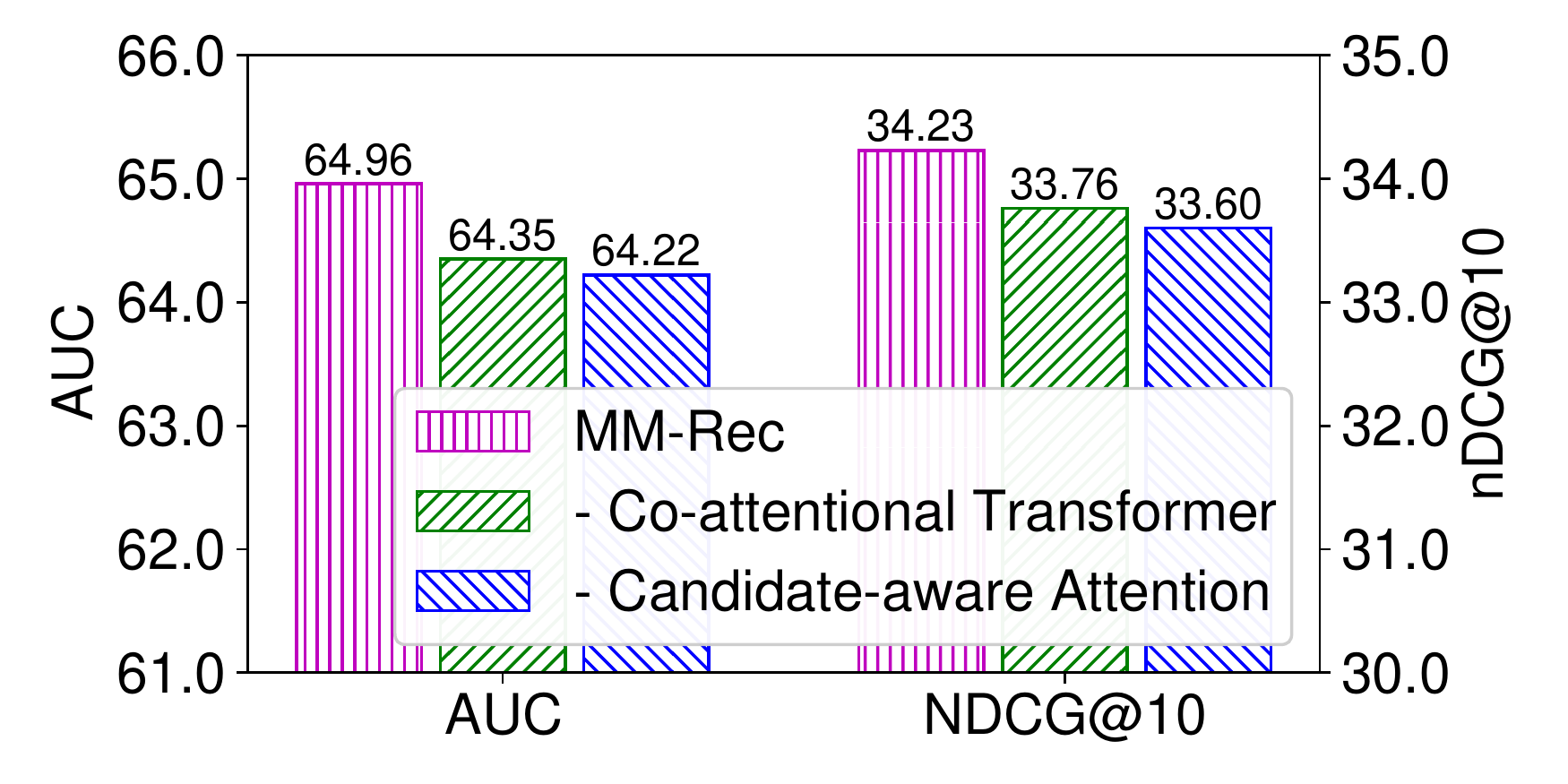}
 \vspace{-0.05in}
\caption{Effect of the co-attentional Transformers in ViLBERT and crossmodal candidate-aware attention.}\label{fig.att}\vspace{-0.15in} 
\end{figure}

\begin{figure}[!t]
	\centering
	\includegraphics[width=0.4\textwidth]{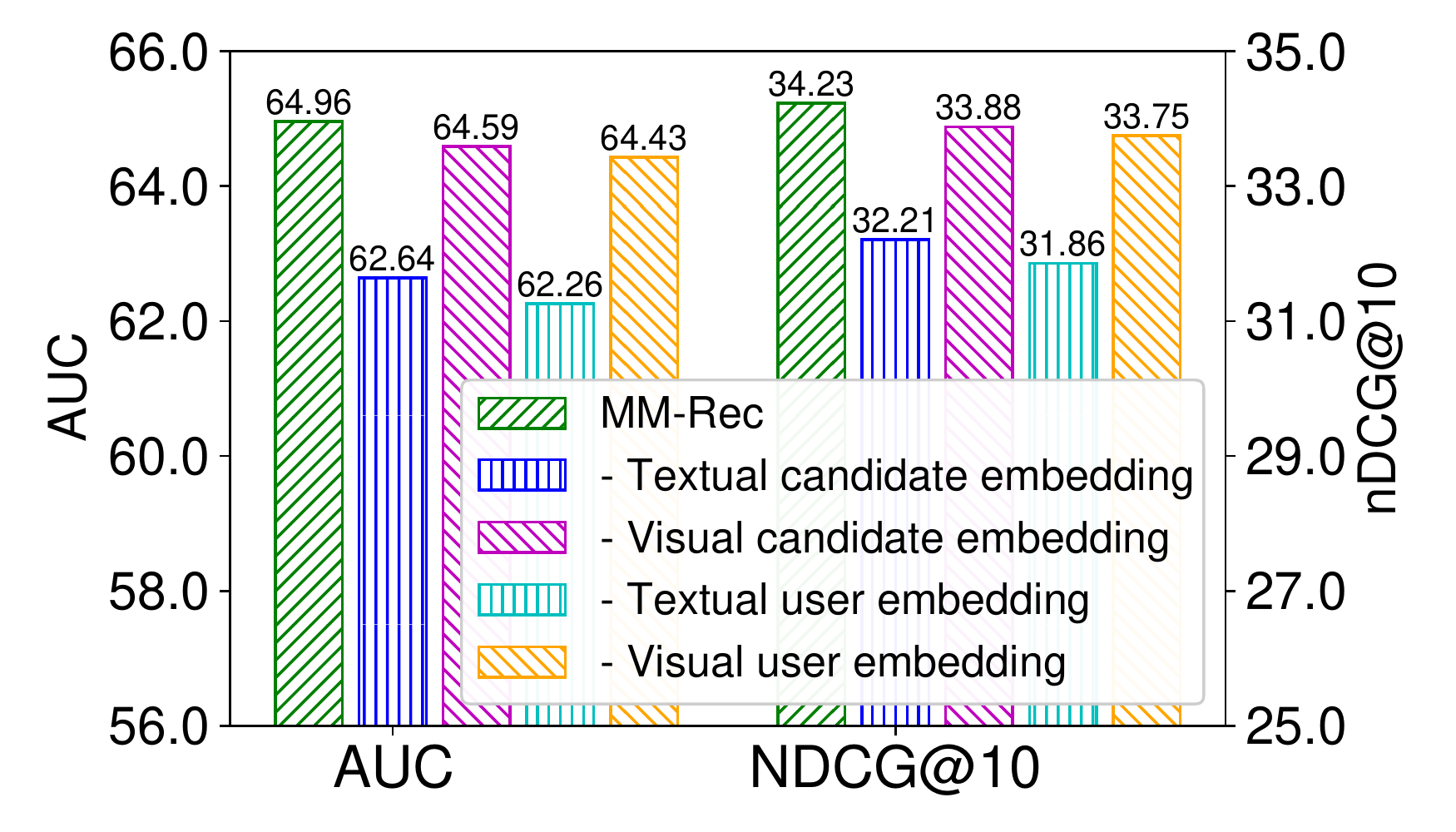}
\vspace{-0.05in}
\caption{Ablation studies on different  embeddings.}\label{fig.emb}\vspace{-0.15in}
\end{figure}

Then we study the effectiveness of the co-attentional Transformers in the ViLBERT model and the crossmodal candidate-aware attention network for user interest modeling.
We compare \textit{MM-Rec} and its variants without co-attentional Transformers or replacing the crossmodal candidate-aware attention with the vanilla attention mechanism used in~\cite{wu2019}.
The results are shown in Fig.~\ref{fig.att}.
We find incorporating co-attentional Transformers network is helpful.
This is because there is inherent relatedness between news title and image in representing the news content and attracting news clicks.
Thus, modeling their interactions can enhance their representations.
In addition, the candidate-aware attention network is useful.
It is because different clicked news usually have different importance for modeling users' specific  interests in candidate news, and selecting them according to their crossmodal relatedness with candidate news can help better match user interest.

\subsection{Ablation Study on Embeddings}

We present several ablation studies on the multimodal candidate news embedding and clicked news embedding for user modeling.
The results are illustrated in Fig.~\ref{fig.emb}.
We find the performance drops when any of the candidate news embeddings or the user embeddings is removed, which shows that all of them are useful.
It shows that both textual and  visual information are useful for news and user modeling.
In addition, textual information plays more important roles in news and user modeling, which is consistent with the results in Fig.~\ref{fig.view}.
This is a very interesting phenomenon because texts are usually less attractive than images.
We think this is mainly because a single news image usually cannot comprehensively summarize news content, and it may  be very challenging to understand visual information accurately.

\subsection{Case Study}

\begin{figure}[!t]
	\centering
	\includegraphics[width=0.9\linewidth]{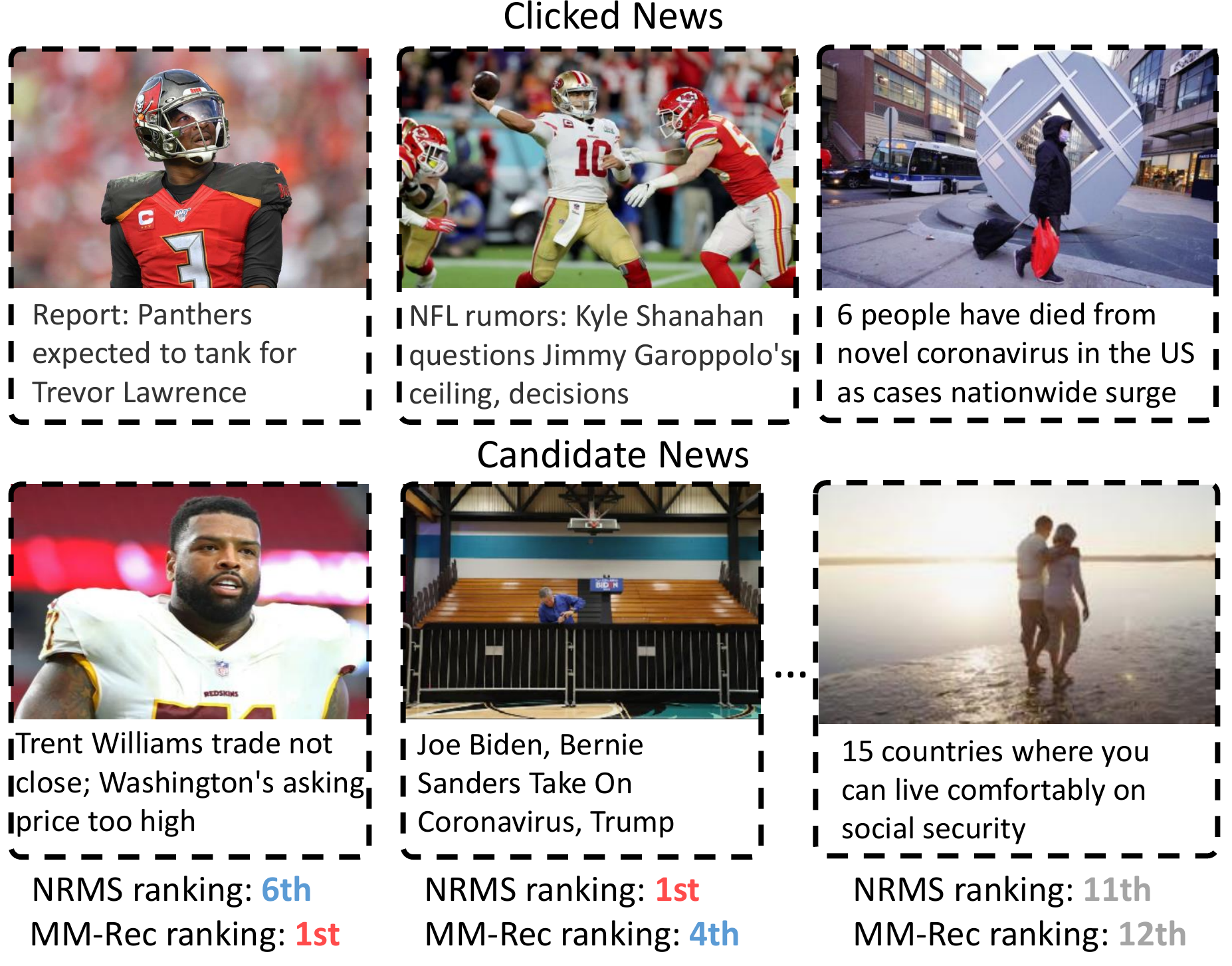}
 
\caption{The clicked news of a user and the rankings of candidate news given by \textit{NRMS} and \textit{MM-Rec}. Only the first candidate news is clicked.}\label{fig.case} \vspace{-0.1in}
\end{figure}

We conduct several case studies to visually demonstrate the effectiveness of incorporating multimodal information into news recommendation.
We show the clicked news of a random user and the rankings given by \textit{NRMS} and \textit{MM-Rec} in Fig.~\ref{fig.case}. 
We find that both \textit{NRMS} and \textit{MM-Rec} assign the last candidate news low rankings, because from its title we can easily infer that it is irrelevant to the user interests.
However,  the \textit{NRMS} model fails to promote the first candidate news, which is highly related to the user's clicked news about NFL.
This may be because it is difficult to measure their relevance solely based on their  titles.
Fortunately, our \textit{MM-Rec} method ranks the first candidate news at the top position because it is easy to match it with user interests based on visual information.
These results show the effectiveness of multimodal information in news recommendation.

\section{Conclusion}\label{sec:Conclusion}

In this paper we present MM-Rec, which can utilize both textual and visual news information to model news for recommendation.
We use a visiolinguistic model to encode both news texts and images and capture their inherent crossmodal relatedness.
In addition, we propose a crossmodal candidate-aware attention network to select relevant clicked news based on their crossmodal relevance to candidate news, which can better model users' specific interest in candidate news.
Experiments show MM-Rec can effectively exploit multimodal news information to improve news recommendation.

\bibliographystyle{ACM-Reference-Format}
\bibliography{main}

\end{document}